\documentclass[twocolumn,showpacs,floatfix]{revtex4}%
\usepackage{graphicx}%
\usepackage{amsmath}%
\usepackage{amsfonts}%
\usepackage{amssymb}
\usepackage{bm}
\usepackage{slashbox}     

\def\s{{\sigma}}
\def\e{{\epsilon}}
\def\k{{ {\bm k} }}

\def\q{{ {\bm q} }}
\def\Q{{ {\bm Q} }}

\def\0{{ {\bm 0} }}
\def\w{{\omega}}
\def\a{{\alpha}}

\def\Tc{{ T_{\rm c} }}
\allowdisplaybreaks[4]

\begin{document}
\title{
Reproduction of Experimental Gap Structure in LiFeAs based on the \\
Orbital-Spin Fluctuation Theory: 
$s_{++}$-wave, $s_\pm$-wave, and hole-$s_\pm$-wave states
}
\author{
Tetsuro \textsc{Saito}$^{1}$, 
Seiichiro \textsc{Onari}$^{2}$,
Youichi \textsc{Yamakawa}$^{1}$, 
Hiroshi \textsc{Kontani}$^{1}$,
Sergey V. Borisenko$^{3}$, and
Volodymyr B. Zabolotnyy$^{3}$
}

\date{\today }

\begin{abstract}

The absence of nesting between electron and hole-pockets 
in LiFeAs with $T_{\rm c}=18$K attracts great attention,
as an important hint to understand the pairing mechanism of
Fe-based superconductors.
Here, we study the five-orbital model of LiFeAs
based on the recently-developed orbital-spin fluctuation theories.
It is found that the experimentally observed gap structure of LiFeAs,
which is a ``fingerprint'' of the pairing mechanism,
is quantitatively reproduced in terms of the orbital-fluctuation-mediated
$s_{++}$-wave state.
Especially, the largest gap observed on the small two hole-pockets 
composed of ($d_{xz}, d_{yz}$) orbitals can be explained, 
and this is a hallmark of the orbital-fluctuation-mediated superconductivity.
The $s_{++}$-wave gap structure becomes more anisotropic
in the presence of weak spin fluctuations.
As the spin fluctuations increase,
we obtain the ``hole-$s_\pm$-wave state'', in which 
only the gap of the large hole-pocket made of $d_{xy}$-orbital is sign-reversed,
due to the cooperation of orbital and spin fluctuations.
This gap structure with ``sign-reversal between hole-pockets''
is similar to that recently reported in (Ba,K)Fe$_2$As$_2$.

\end{abstract}

\address{
$^1$ Department of Physics, Nagoya University,
Furo-cho, Nagoya 464-8602, Japan. 
\\
$^2$ Department of Applied Physics, Nagoya University,
Furo-cho, Nagoya 464-8603, Japan.
\\
$^3$ Leibniz-Institute for Solid State Research, IFW-Dresden, 
D-01171 Dresden, Germany  
}
 

\sloppy

\maketitle


\section{Introduction}
\label{sec:Intro}
The most remarkable feature of Fe-based superconductors
would be the amazing variety of the normal and superconducting states
in various materials.
The high-$\Tc$ state with $\Tc>30$K is realized by 
electron-doping, hole-doping, in addition to isovalent-doping, 
irrespective of the huge change of the Fermi surfaces (FSs).
In Ba122 systems, the superconducting phase is next to the 
orthorhombic ($C_2$) structure phase, accompanied by the magnetic order.
Near the structural and magnetic quantum critical points,
strong orbital and spin fluctuations are observed
\cite{NMR-Fujiwara,NMR-Ishida,NMR-Zheng,C66-Fer,C66-Yoshizawa,C66-Bohmer,Raman},
and these fluctuations would be the origin of superconductivity.
In FeSe$_x$Te$_{1-x}$, in contrast, the optimum $T_{\rm c}$ 
is realized near the structural quantum-critical-point 
at $x\approx0.6$ \cite{FeSe1},
whereas magnetic order is absent for $x>0.4$.
In heavily H-doped La1111 and (La,P) co-doped Ca122,
high-$\Tc$ above 40K is realized near the isostructural ($C_4$) 
transition phase.

Also, the shape of the FSs, which is essential for the electronic properties
of each material, is strongly material-dependent
\cite{Borisenko-various}:
In La1111, the band structure is almost two-dimensional
and the FSs are mainly composed of the $d_{xz}$, $d_{yz}$ and $d_{xy}$ orbitals.
In Ba122, the band structure has three-dimensional character,
and the $d_{3z^2-r^2}$ orbital also contributes to the FS.
In FeSe$_x$Te$_{1-x}$, the Fermi momentum $k_{\rm F}$ is less than
one-fifth of that of Ba122, and the Fermi energy is just $\sim 100$K
\cite{Borisenko-FeSe}.
In heavily H-doped La1111 and (La,P) co-doped Ca122,
high-$\Tc$ ($\gtrsim40$K) is realized 
irrespective that the hole-pockets are very tiny or absent.
These experimental facts strongly indicate the wide variety of 
the pairing mechanisms in Fe-based superconductors,
and quantitative analysis based on the realistic tight-binding model 
is required for each compound.

Up to now, the spin-fluctuation-mediated $s_\pm$-wave state
with sign-reversal between hole- and electron-pockets
had been studied by many authors
\cite{Kuroki,Mazin,Chubukov,Hirschfeld}.
In La1111 systems, however,
the relation between the strength of spin fluctuations and $\Tc$ is less clear.
For example, $\Tc$ of 14\% F-doped LaFeAsO increases
from 23K to 43K by applying the pressure 3GPa,
although the $1/T_1$ remains small and almost unchanged.
Later, the orbital-fluctuation-mediated $s_{++}$-wave state
without sign-reversal had been proposed
\cite{Kontani-RPA,Onari-SCVC}.
The robustness of $T_{\rm c}$ against the randomness in various Fe-pnictides
\cite{Sato-imp,Li-imp,Nakajima-imp}
is consistent with the $s_{++}$-wave state 
\cite{Onari-imp,Yamakawa-imp}.
Also, resonance-like peak in the neutron scattering spectrum 
is explained in term of the $s_{++}$-wave state
by taking the realistic inelastic scattering into account 
\cite{Onari-neutron}.

\begin{figure}[!htb]
\includegraphics[width=0.9\linewidth]{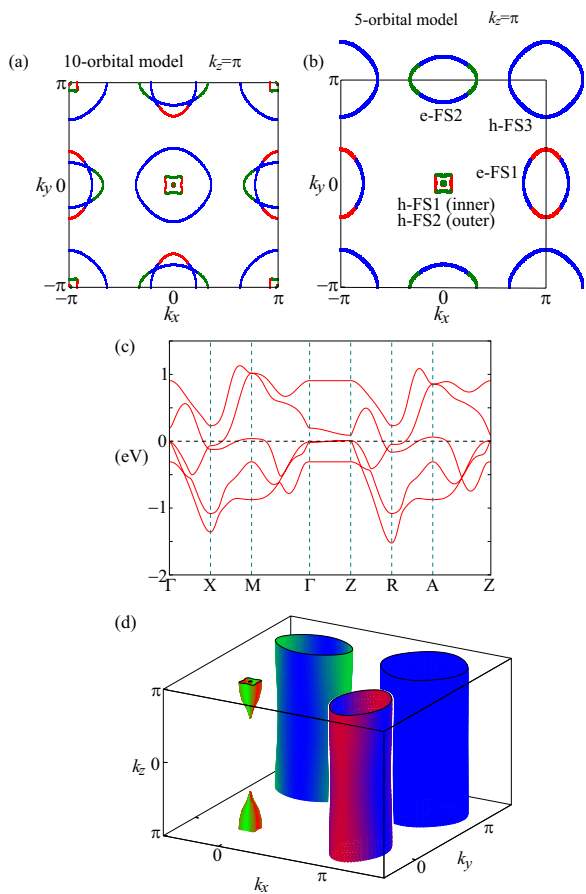}
\caption{
(Color online)
The FSs in the $k_z=\pi$ plane
of the three-dimensional ten-orbital model (a) and five-orbital model (b) 
for LiFeAs. The green, red and blue lines correspond to
$xz$, $yz$ and $xy$ orbitals, respectively.
In (b), h-FS1,2,3 are hole-like, and e-FS1,2 are electron-like.
(c) The dispersion of the band structure of the five-orbital model.
(d) The three-dimensional shape of the FSs of the five-orbital model.
}
\label{fig:FS}
\end{figure}

In the study of the pairing mechanism, the detailed gap structure 
given by the angle-resolved-photoemission-spectroscopy (ARPES)
offers us very useful information.
For this purpose, LiFeAs ($\Tc=18$K) is favorable since 
very clean single crystals can be synthesized.
For this reason, 
the intrinsic gap structure free from the impurity effect
can be obtained in the case of LiFeAs.
The detailed gap structure of LiFeAs 
had been obtained by ARPES \cite{Borisenko-LiFeAs,Ding-LiFeAs}.
The FSs given in Ref. \cite{Borisenko-LiFeAs}
are shown in Fig. \ref{fig:FS} (a), which are reproduced by 
the ten-orbital tight-binding model (two-Fe unit cell).
Figure  \ref{fig:FS} (b) shows the FSs in the five-orbital model 
(single Fe unit cell) obtained by unfolding the original ten-orbital model.
Both models are equivalent mathematically, and the unfolding is performed 
by following the procedure in Ref. \cite{Miyake}.

The bad nesting in LiFeAs between hole-like FSs (h-FSs) 
and electron-like FSs (e-FSs) attracts great attention,
as an important hint to understand the variety and commonness of 
the pairing mechanism in Fe-based superconductors.
Consistently, the observed spin fluctuations are moderate
according to NMR measurements \cite{NMR-LiFeAs}
and neutron scattering measurements 
\cite{neutron-LiFeAs1,neutron-LiFeAs2,neutron-LiFeAs3}.

In Ref. \cite{Hirschfeld-LiFeAs},
the spin fluctuation mediated $s_\pm$-wave state had been studied
based on the ten-orbital model for LiFeAs.
The obtained gap functions on the tiny hole-pockets
h-FS1 and h-FS2 in Fig. \ref{fig:FS} (b) are very small 
when the filling of electrons per Fe-site is $n=6.0$,
although they are the largest in the ARPES measurement 
\cite{Borisenko-LiFeAs,Ding-LiFeAs}
and the Scanning Tunneling Microscopy (STM) measurement \cite{QPI}.
Thus, it is an important challenge to verify 
to what extent the experimental gap structure is reproduced 
based on the orbital fluctuation theories.

In this paper,
we study the five-orbital model of LiFeAs
based on the recently-developed orbital-spin fluctuation theories
\cite{Kontani-RPA,Onari-SCVC}.
When only the orbital fluctuations develop,
the anisotropic $s_{++}$-wave state without sign-reversal is obtained.
In this case, experimentally observed gap structure of LiFeAs,
especially the largest gap experimentally observed on h-FS1 and h-FS2,
is quantitatively reproduced.
This is a hallmark of the orbital-fluctuation-mediated superconductivity
since the spin fluctuation scenario 
predicts the smallest gap on h-FS1 and h-FS2.
When orbital and spin fluctuations coexist,
we can obtain the ``hole-$s_\pm$-wave state'', in which 
the gap structure with ``sign-reversal between hole-pockets''is realized.
This exotic gap structure 
had been discussed in (Ba,K)Fe$_2$As$_2$ experimentally,
\cite{Watanabe-shpm,Ding-shpm}
and it might be realized in other Fe-based superconductors.

We stress that the experimental gap structure
is a ``fingerprint'' of the pairing mechanism.
For example, in Ba122 and Sr122, the ``horizontal-node''
on the $d_{3z^2-r^2}$-orbital h-FS around the Z point
was predicted by spin fluctuation theories
\cite{Suzuki,Saito-loop},
since $d_{3z^2-r^2}$-orbital is absent on the e-FSs.
In contrast, the horizontal-node is absent in the 
orbital-fluctuation-mediated $s$-wave state,
due to the strong inter-orbital fluctuations \cite{Saito-loop}.
The latter is supported by recent ARPES measurements
for optimal BaFe$_2$(As,P)$_2$ \cite{Shimo-Science,Yoshida} 
and in-plane field angle dependence of the thermal conductivity 
\cite{Yamashita}.

\section{Formalism}
\label{sec:Formalism}

In this paper, we set $x$ and $y$ axes parallel to the nearest Fe-Fe bonds,
and the orbital $z^2$, $xz$, $yz$, $xy$, and $x^2-y^2$ are denoted as
1, 2, 3, 4, and 5 respectively.
The three-dimensional ten-orbital tight-binding model
had been obtained in Ref. \cite{Hirschfeld-LiFeAs}
by fitting the experimentally observed dispersion
reported in Ref. \cite{Borisenko-LiFeAs}, and its 
FSs are shown in Fig. \ref{fig:FS} (a).
In this model, the band renormalization due to 
the mass enhancement $m^*/m_b \sim2$ is taken into account.
To simplify the numerical calculation,
we derive the five-orbital model by unfolding the 
original ten-orbital model \cite{Miyake}.
The FSs and the band dispersion of the five-orbital model 
are shown in Fig. \ref{fig:FS} (b) and (c), respectively.
The three-dimensional FSs of the five-orbital model are show
in Fig. \ref{fig:FS} (d).

The kinetic term of the five-orbital model is given as
\begin{eqnarray}
\hat{H}^{0}&=& \sum_{a b l m \sigma} t_{l,m}(\bm{R}_{a}-\bm{R}_{b}) 
c^{\dagger}_{l \s}(\bm{R}_{a}) c_{m \s} (\bm{R}_{b})
 \nonumber \\
&=& \sum_{\k l m \s} \left\{ \sum_{a} t_{l,m}(\bm{R}_{a}) 
e^{i \bm{k} \cdot \bm{R}_{a}} \right\} c^{\dagger}_{l \s}(\bm{k}) c_{m \s} (\bm{k}) ,
\label{eqn:H0}
\end{eqnarray}
where $a,b$ represent the Fe-sites, 
$l,m = 1-5$ represent the $d$ orbital, and $\s = \pm 1$ is the spin index.
$\bm{R}_{a}$ is the position of Fe-site,
$c^{\dagger}_{l \s}(\bm{R}_{a})$ is the creation operator of the $d$ electron,
and $t^{l,m}(\bm{R}_{a})$ is the hopping integral.
The values of $t^{l,m}_{a,b}$ are shown in Appendix \ref{sec:Appendix-TB}.
Figure \ref{fig:DOS} (a) and (b) show the 
inverse of the Fermi velocity on the $i$-th FS, $1/v_{\rm F}^i(\k)$,
in $k_z=0$ and $k_z=\pi$ planes, respectively.
The horizontal axis is $\theta={\rm tan}^{-1}(\bar{k}_y/\bar{k}_x)$,
where $(\bar{k}_x,\bar{k}_y)$ is the momentum on the FS
with the origin at the center of each pocket.
Figure \ref{fig:DOS} (c)-(f) show the $l$-orbital weight on the $i$-th FS,
given by $|U_{l,i}(\k)|^2=|\langle \k,l|\k,i\rangle|^2$ at the Fermi momentum.

\begin{figure}[!htb]
\includegraphics[width=0.99\linewidth]{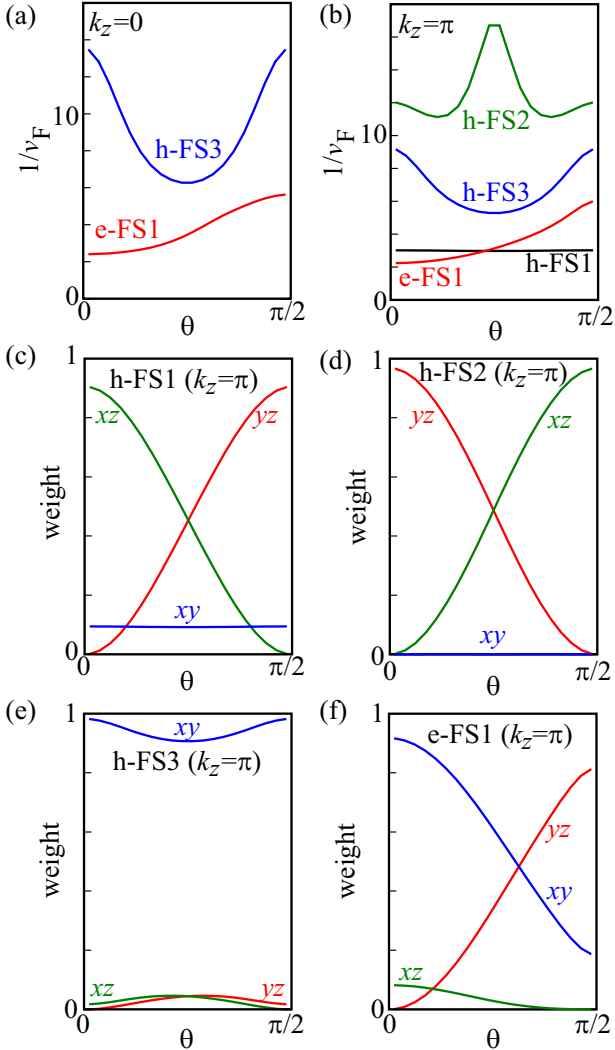}
\caption{
(Color online)
(a,b) Inverse of the Fermi velocity on the $i$-th FS $1/v_{\rm F}^i(\k)$.
The horizontal axis is $\theta={\rm tan}^{-1}(k_y/k_x)$.
(c-f) The weight of each $d$-orbital on the $i$-th FS.
}
\label{fig:DOS}
\end{figure}

Next, we explain the interaction term.
We introduce both the Coulomb interaction ($U$, $U'$, $J=(U-U')/2$) 
and quadrupole interaction.
The latter are induced by the electron-phonon (e-ph) interaction
due to Fe ion oscillations as follows, \cite{Kontani-RPA}
\begin{eqnarray}
V_{\rm{quad}}=- g(\w_l) \sum_a^{\rm site} \left( 
{\hat O}^a_{yz} \cdot{\hat O}^a_{yz} + {\hat O}^a_{xz} 
\cdot{\hat O}^a_{xz} \right) ,
 \label{eqn:Hint}
\end{eqnarray}
where $g(\w_l)=g \cdot \w_0^2/(\w_l^2+\w_0^2)$, and 
$g=g(0)$ is the quadrupole interaction at $\omega_l=0$.
$\omega_0$ is the cutoff energy of the quadrupole interaction.
$\hat{O}_{\Gamma}^a=\sum_{l,m}o^{l,m}_\Gamma {\hat m}_{l,m}^a$ 
(${\hat m}_{l,m}^a= \sum_\s c_{l\s}^\dagger(\bm{R}_a) c_{m\s}(\bm{R}_a)$)
is the quadrupole operator at site $\bm{R}_a$ introduced 
in Ref. \cite{Kontani-RPA}:
The non-zero coefficients of $o^{l,m}_\Gamma=o^{m,l}_\Gamma$ are 
$o_{xz}^{2,5}=o_{xz}^{3,4}=\sqrt{3}o_{xz}^{1,2}=1$, and
$-o_{yz}^{3,5}=o_{yz}^{2,4}=\sqrt{3}o_{yz}^{1,3}=1$.
Thus, $\hat{V}_{\mathrm{quad}}$ has many non-zero inter-orbital elements.
As explained in Ref. \cite{Kontani-RPA},
$g$ is induced by in-plane Fe-ion oscillations.
Also, the Aslamazov-Larkin type vertex correction (AL-VC)
due to Coulomb interaction produces large effective 
quadrupole interaction $g$ \cite{Onari-SCVC}.
Thus, the quadrupole interaction in eq. (\ref{eqn:Hint}) 
is derived from both mechanisms.

Now, we perform the RPA for the present model,
by using $64 \times 64 \times 16 \ \bm{k}$ meshes.
We fix the temperature at $T=0.01$ eV, 
and set the filling of each Fe-site as $n=6.0$.
Hereafter, the unit of energy as eV.
The irreducible susceptibility in the five-orbital model is given by
\begin{eqnarray} 
\chi^{(0)}_{ll',mm'} \left( q \right) =
- \frac{T}{N} \sum_k G_{l,m}^{(0)} \left( k+ q \right) G_{m',l'}^{(0)} \left( k \right),
\label{eqn:chi0}
\end{eqnarray}
where $q = ( \bm{q}, \omega_l )$ and $k=( \bm{k} , \epsilon_n)$.
$\epsilon_n = (2n + 1) \pi T$ and $\w_l = 2l \pi T$ 
are the fermion and boson Matsubara frequencies.
$\hat{G}^{(0)} ( k ) = [ i \epsilon_n + \mu - \hat{h}^{(0)}_{\bm{k}} ]^{-1}$
is the Green function in the orbital basis,
where $\hat{h}^{(0)}_{\bm{k}}$ is the matrix representation of
$\hat{H}^{(0)}$ and $\mu$ is the chemical potential.
In the RPA, the susceptibilities for spin and charge sectors are given by 
\cite{Takimoto}
\begin{eqnarray} 
&&\hat{\chi}^{\mathrm{s}} \left( q \right) = \frac{\hat{\chi}^{(0)} \left( q \right)}{\hat{1} - \hat{\Gamma}^{\mathrm{s}} \hat{\chi}^{(0)} \left( q \right)}, 
\label{eqn:chis} \\
&&\hat{\chi}^{\mathrm{c}} \left( q \right) = \frac{\hat{\chi}^{(0)} \left( q \right)}{\hat{1} - \hat{\Gamma}_g^{\mathrm{c}} (\omega_l) \hat{\chi}^{(0)} \left( q \right)},
\label{eqn:chic}
\end{eqnarray}
where
\begin{equation}
(\Gamma^{\mathrm{s}})_{l_{1}l_{2},l_{3}l_{4}} = \begin{cases}
U, & l_1=l_2=l_3=l_4 \\
U' , & l_1=l_3 \neq l_2=l_4 \\
J, & l_1=l_2 \neq l_3=l_4 \\
J' , & l_1=l_4 \neq l_2=l_3 \\
0 , & \mathrm{otherwise}
\end{cases}
\end{equation}
\begin{equation}
\hat{\Gamma}_g^{\mathrm{c}} ( \omega_l )= \hat{\Gamma}^{\mathrm{c}} - 2\hat{V}_{\mathrm{quad}}( \w_l ),
\label{eqn:Gc}
\end{equation}
\begin{equation}
({\hat \Gamma}^{\mathrm{c}})_{l_{1}l_{2},l_{3}l_{4}} = \begin{cases}
-U, & l_1=l_2=l_3=l_4 \\
U'-2J , & l_1=l_3 \neq l_2=l_4 \\
-2U' + J , & l_1=l_2 \neq l_3=l_4 \\
-J' , &l_1=l_4 \neq l_2=l_3 \\
0 . & \mathrm{otherwise}
\end{cases}
\end{equation}

In the RPA, the enhancement of the spin susceptibility $\hat{\chi}^{\mathrm{s}}$
is mainly caused by the intra-orbital Coulomb interaction $U$,
using the ``intra-orbital nesting'' of the FSs.
On the other hand, the enhancement of $\hat{\chi}^{\mathrm{c}}$ in the present model
is caused by the quadrupole-quadrupole interaction in eq. (\ref{eqn:Hint}),
utilizing the ``inter-orbital nesting'' of the FSs.
The magnetic (orbital) order is realized 
when the spin (charge) Stoner factor $\alpha_{\mathrm{s}}$ 
($\alpha_{\mathrm{c}}$),
which is the maximum eigenvalue of 
$\hat{\Gamma}^{\mathrm{s}} \hat{\chi}^{(0)} ( \bm{q} , 0)$, 
($\hat{\Gamma}_g^{\mathrm{c}}(0) \hat{\chi}^{(0)} ( \bm{q} , 0)$), is unity.
Here, the critical value of $U$ is $U_{\mathrm{cr}}=0.448$ eV,
and the critical value of $g$ is $g_{\mathrm{cr}}= 0.132$ eV for $U=0$.
(We note again that the band renormalization due to 
the mass enhancement $m^*/m_b \sim2$ is taken into account
in the present tight-binding model.)

\begin{figure}[!htb]
\includegraphics[width=0.9\linewidth]{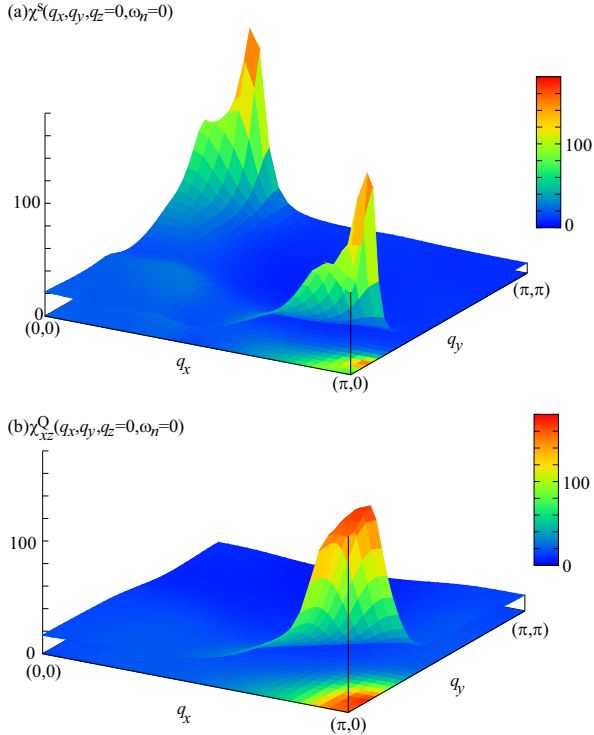}
\caption{
(Color online)
(a) Obtained spin susceptibility
$\chi^{s}(\bm{q},0)$ for $U=0.98U_{\rm cr}$ and $g=0$.
The spin fluctuations develop mainly on the $d_{xy}$-orbital.
(b) Obtained $O_{xz}$-channel quadrupole susceptibility
$\chi_{xz}^{Q}(\bm{q},0)$ for $U=0$ and $g=0.98g_{\rm cr}$,
developed among the $d_{xz}/d_{yz}$-orbitals.
}
\label{fig:chi}
\end{figure}

Figure \ref{fig:chi} (a) shows the obtained spin susceptibility
$\chi^s(\q,0) \equiv \sum_{l,m}\chi^s_{l,l:m,m}(\q,0)$ in the $q_z=0$ plane
given by the RPA for $U=0.439$ and $g=0$.
The spin Stoner factor is $\a_{\rm s}=0.98$.
At $T=0.01$, the obtained peak is incommensurate at $(\pi,\delta)$
with $\delta\approx0.1\pi$,
consistently with the recent neutron scattering experiment 
\cite{neutron-LiFeAs2}.
The relation $\chi^s_{4,4;4,4}(\q,0)\gg \chi^s_{2,2;2,2}(\q,0),\chi^s_{3,3;3,3}(\q,0)$
holds in the present model, due to the intra $d_{xy}$-orbital nesting
between h-FS3 and e-FS.
That is, the spin fluctuations develop mainly on the $d_{xy}$-orbital.

Figure \ref{fig:chi} (b) shows the quadrupole susceptibility
$\chi^Q_\Gamma(\q,0)=\sum_{l,l',m,m'}o_\Gamma^{l,l'}\chi^c_{l,l':m,m'}(\q,0) o_\Gamma^{m',m}$
for the channel $\Gamma=xz$ in the $q_z=0$ plane.
The charge Stoner factor is $\a_{\rm c}=0.98$.
In this model, both $\chi^Q_{xz}(\q,0)$ and $\chi^Q_{yz}(\q,0)$
are the most divergent channels.
For $\Gamma=xz$, the dominant contribution comes from
$\chi^c_{3,4;4,3}(\q,0)\approx \chi^c_{3,4;3,4}(\q,0)$,
due to the inter-orbital nesting (orbital 3 and 4)
between h-FS1,2 and e-FS1.
The obtained $\chi^Q_{xz}(\q,0)$ shows broad peak around
$(\pi,\delta)$ with $|\delta|\lesssim 0.2\pi$.

We note that both $\chi^Q_{xz}(\q,0)$ and $\chi^s(\q,0)$ 
are almost independent of $q_z$.
That is, both the orbital and spin fluctuations are almost two-dimensional.

Next, we solve the linearized Eliashberg equation 
based on the three-dimensional model of LiFeAs.
In order to obtain the fine momentum dependence of the SC gap,
we concentrate on the gap functions only on the FSs
as done in Ref. \cite{Scalapino}:
We used $80 \times 16 \ \bm{k}$ points for each Fermi surface sheet.
Without impurities,
the linearized Eliashberg equation is given as \cite{Scalapino}
\begin{eqnarray}
\lambda_{\mathrm{E}} \Delta^{i} ( \bm{k} , \e_n ) 
&=& \frac{\pi T}{(2 \pi )^3} \sum_{\e_m} \sum_{j}^{\mathrm{FS}} \int_{\mathrm{FS}j} 
\frac{d \bm{k}'_{\mathrm{FS}j}}{v^{j} ( \bm{k}')}
\nonumber \\
& &\times V^{ij}( \bm{k} , \bm{k}' , \e_n - \e_m ) 
\frac{\Delta^{j}( \bm{k}' , \e_m )}{| \e_m |} ,
\label{eqn:Eliasheq}
\end{eqnarray}
where $\lambda_{\mathrm{E}}$ is the eigenvalue that reaches unity at $T=T_{\mathrm{c}}$.
$i$ and $j$ denote the FSs, and 
$ \Delta^{i} ( \bm{k} , \e_n )$ is the 
gap function on the $i$-th FS at the Fermi momentum $\bm{k}$.
The integral in eq. (\ref{eqn:Eliasheq}) means the surface integral 
on the $j$-th FS.
The paring interaction $V$ in eq. (\ref{eqn:Eliasheq}) is
\begin{multline}
V^{i j} ( \bm{k} , \bm{k}' , \e_n - \e_m) = \sum_{l_i} U_{l_1 ,i}^{*} ( \bm{k} ) U_{l_4 , i} ( \bm{k} ) \\
\times V_{l_1 l_2,l_3 l_4} (\bm{k} - \bm{k}' , \e_n - \e_m )
U_{l_2, j} ( \bm{k}' ) U_{l_3, j}^{*} ( \bm{k}' ),
\label{eqn:Vint}
\end{multline}
\begin{gather}
\hat{V} = \hat{V}^{\mathrm{c}} + \hat{V}^{\mathrm{s}} + \hat{V}^{(0)}
, \label{eqn:Worb} \\
\hat{V}^{\mathrm{c}} = \frac{1}{2} \hat{\Gamma}_g^{\mathrm{c}} \hat{\chi}^{\mathrm{c}} \hat{\Gamma}_g^{\mathrm{c}}, \ \ \ 
\hat{V}^{\mathrm{s}} = -\frac{3}{2} \hat{\Gamma}^{\mathrm{s}} \hat{\chi}^{\mathrm{s}} \hat{\Gamma}^{\mathrm{s}} ,\label{eqn:Worb-cs} \\
\hat{V}^{(0)} = \frac{1}{2} ( \hat{\Gamma}_g^{\mathrm{c}}-\hat{\Gamma}^{\mathrm{s}}) ,
\end{gather}
where $U_{l, i} ( \bm{k} ) = \langle \bm{k} ; l | \bm{k} ; i \rangle$
is the transformation unitary matrix
between the band and the orbital representations.

In this calculation, we simplify the energy dependence of $\hat{V}$.
We assume that $\hat{V}^{\xi}$ ($\xi=\mathrm{c},\mathrm{s}$) can be separated into
the momentum and orbital dependent part $\hat{V}^{\xi} (\bm{k} , \w_l = 0)$
and energy dependent part $g_{\xi}(\w_l )$:
\begin{equation}
\hat{V}^{\xi} (\bm{k} , \w_l) = \hat{V}^{\xi}(\bm{k}, \w_{l} = 0) \times g_{\xi}(\w_l).
\end{equation}
We calculated $\hat{V}^{\xi}(\bm{k}, \w_{l} = 0)$ without approximation.
On the other hand, $g_{\xi}(\w_l )$ is determined as
\begin{equation}
g_{\xi}(\w_l)
= \mathrm{Re} \left[
\frac{V^{\xi}_{\mathrm{max}} (\w_l)}
{V^{\xi}_{\mathrm{max}} ( \w_l = 0) } \right] ,
\end{equation}
where $V^{\xi}_{\mathrm{max}} (0)$ is the largest value of
$V^{\xi}_{l_1 l_2 , l_3 l_4} ( \bm{k}, \w_l = 0)$
for any $\{l_i\}$ and $\bm{k}$.
It is verified that 
this simplification affects the momentum dependence of
the SC gap functions only quantitatively, although 
the obtained $\lambda_{\mathrm{E}}$ is slightly underestimated.
Thus, this approximation would be appropriate for the 
present purpose, that is, the analysis of the anisotropy of the SC gap.

\section{Superconducting Gap Functions}
\label{sec:RPA}

In this section, we analyze the linearized Eliashberg equation,
eq.(\ref{eqn:Eliasheq}),
and obtain the three-dimensional gap function $\Delta^i(\theta,k_z)$,
defined on the Fermi surface sheet $i$.
Here, we divide the valiables $\theta=[0,2\pi]$ and $k_z=[-\pi,\pi]$ 
into $80$ and $16$ meshes, respectively,
and use 512 Matsubara frequencies.
The pairing interaction in eq. (\ref{eqn:Vint})
is given by the RPA,
assuming that $J=J'$ and $U=U'+2J$, and fix the ratio $J/U=1/6$.
The used parameters are $T=0.01$ and $\omega_{0}= 0.02$.

\subsection{Orbital-fluctuation-mediated $S_{++}$-wave state}

\begin{figure}[!htb]
\includegraphics[width=0.9\linewidth]{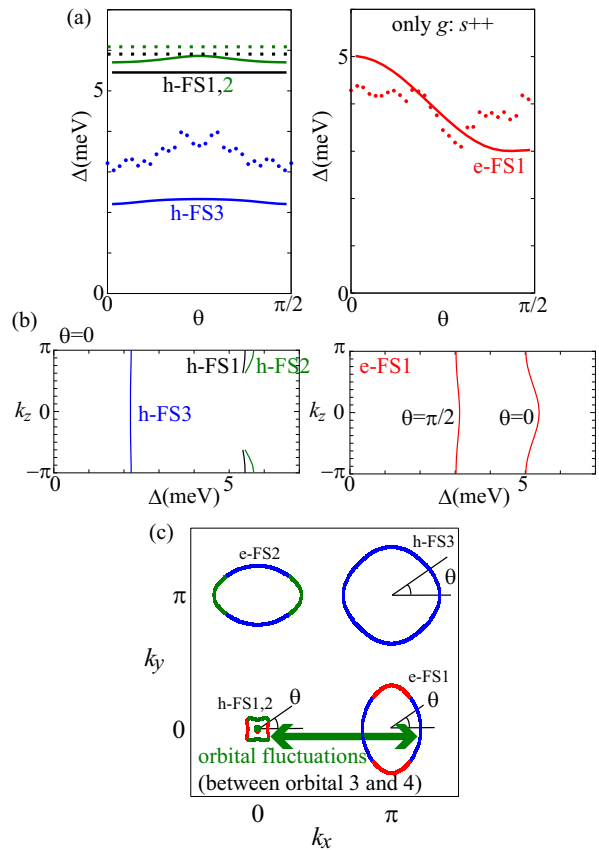}
\caption{
(Color online)
(a) Obtained $s_{++}$-wave gap functions 
for $U=0$ and $g=0.129$ in the $k_z=\pi$-plane. 
The eigenvalue is $\lambda_E=0.64$.
The dotted lines represent the experimental data
given by the ARPES measurement in Ref. \cite{Borisenko-LiFeAs}.
(b) $k_z$-dependence of the gap functions.
(c) Explanation for the orbital dependence of the 
gap functions due to orbital fluctuations.
}
\label{fig:gap1}
\end{figure}

We first discuss the $s_{++}$-wave state realized by 
orbital fluctuations:
Figure \ref{fig:gap1} (a) shows the 
obtained gap functions in the case of $g=0.129$ and $U=0$ ($\a_{\rm c}=0.98$)
in the $k_z=\pi$-plane.
As for the hole-pockets,
the gap functions on the h-FS1,2 composed of ($d_{xz},d_{yz}$)-orbitals
are the largest, while the gap on the h-FS3 
composed of $d_{xy}$-orbital is the smallest.
These results are quantitatively consistent with the 
experimental data \cite{Borisenko-LiFeAs} shown in dotted lines.
(We adjust the magnitude of gap functions since it cannot be determined
by solving the linearized gap equation.)

As for the electron-pockets,
the gap function has the local maxima at $\theta=0$,
and the minimum point is $\theta\approx 0.4\pi$.
This result is also consistent with the 
experimental data \cite{Borisenko-LiFeAs}.
In Appendix \ref{sec:Appendix-GAP},
we show the $s_{++}$-wave gap for smaller $g$ ($\a_{\rm c}=0.90$),
and find that the gap structure is essentially independent 
of the strength of orbital fluctuations.
Therefore, overall experimental data
are quantitatively reproduced by the orbital fluctuation theory.
In Fig.  \ref{fig:gap1} (b), we show the three-dimensional gap structure.
The gap function on each FS is almost independent of $k_z$.
Note that h-FS1 and h-FS2 appear only for $k_z\approx\pm\pi$; 
see Fig. \ref{fig:FS} (d).

In Fig.  \ref{fig:gap1} (c), we discuss the origin of 
the orbital- and FS-dependences of the gap functions:
The broad peak of the quadruple susceptibility $\chi^Q_{xz}(\q,0)$ at 
$\q\approx(\pi,\delta)$ with $|\delta|\lesssim0.2\pi$
in Fig. \ref{fig:chi} (b) is mainly given by the 
inter-orbital nesting between h-FS1,2 (orbital 2,3) and e-FS1 (orbital 4).
For this reason, the maximum gap is realized on h-FS1 ($\Delta^{\rm h}_1$), 
h-FS2 ($\Delta^{\rm h}_2$), and e-FS1 ($\Delta^{\rm e}_1$) at $\theta=0$.
The gap size of h-FS3 ($\Delta^{\rm h}_3$) is the smallest, and its
maximum is located at $\theta=\pi/4$,
Therefore, the experimentally observed gap functions
are  understood based on the orbital fluctuation theory very well.

\subsection{Spin-fluctuation-mediated $S_\pm$-wave state}

\begin{figure}[!htb]
\includegraphics[width=0.9\linewidth]{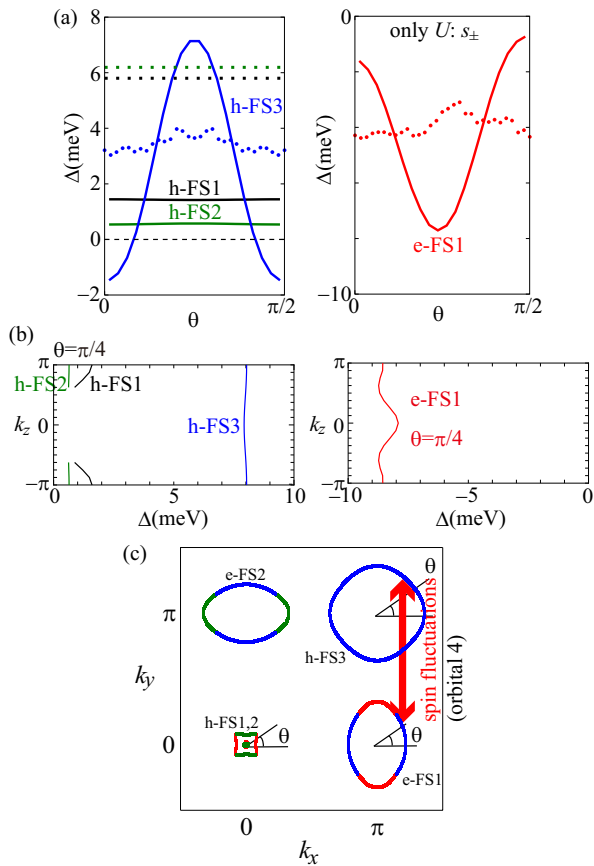}
\caption{
(Color online)
(a) Obtained $s_\pm$-wave gap functions
for $U=0.439$ and $g=0$ in the $k_z=\pi$-plane.
The eigenvalue is $\lambda_E=0.79$.
(b) $k_z$-dependence of the gap functions.
(c) Explanation for the orbital dependence of the 
gap functions due to spin fluctuations.
}
\label{fig:gap2}
\end{figure}

Next, we discuss the $s_\pm$-wave state realized by spin fluctuations:
Figure \ref{fig:gap2} (a) shows the 
obtained gap structure in the case of $g=0$ and $U=0.439$ ($\a_{\rm s}=0.98$)
in the $k_z=\pi$-plane. 
The gap functions are almost independent of $k_z$,
except that h-FS1,2 exist only for $k_z\sim\pi$.
The obtained gap structure is essentially independent
even if smaller $U$ is used.
Similarly to the previous study in Ref. \cite{Hirschfeld-LiFeAs},
the gap functions on the h-FS1,2 are very small.
However, this result is opposite to the experimental data
shown by dotted lines.
The $k_z$-dependence of the gap functions for $\theta=\pi/4$
are shown in Fig.  \ref{fig:gap2} (b).
All gaps depend on $k_z$ only slightly.

In addition, the obtained $\theta$-dependence of the gap on the e-FS1 is  
very different from the experimental data.
Both  $\Delta^{\rm h}_3$ and $\Delta^{\rm e}_1$ show the 
maximum values at $\theta\approx\pi/4$, because of the reason that 
they are connected by the wavevector of the
spin fluctuations $\Q\approx(\pi,0),(0,\pi)$ shown in Fig. \ref{fig:gap2} (c).
In addition, the gap function of h-FS3 has eight nodes 
inconsistently with experiments.
We verified these eight nodes disappear by using larger value of $J/U\sim0.4$
($U'=U-2J\sim0.2U$) as used in Ref. \cite{Hirschfeld-LiFeAs}.

In Appendix \ref{sec:Appendix-GAP},
we show the $s_\pm$-wave gap for smaller $U$ ($\a_{\rm s}=0.90$).
In this case, the magnitude of $\Delta^{\rm h}_{1,2}$ becomes
relatively large.
On the other hand, the nodal gap appears on the e-FSs, 
inconsistently with experiments.
Thus, the overall experimental data
is difficult to be explained by the spin fluctuation theory.



\subsection{Coexistence of orbital and spin fluctuations:
$s_{++}$-wave and hole-$s_\pm$-wave states}

Now, we discuss the superconducting state
when the orbital and spin fluctuations coexist.
In the case of BaFe$_2$(As,P)$_2$,
the coexistence of both fluctuations 
produces the three-dimensional loop-shape nodes on 
electron-like FSs, as discussed in Ref. \cite{Saito-loop}.
In the present model for LiFeAs, we find that the
coexistence of orbital and spin fluctuations 
leads to a very exotic $s$-wave state, since the band structure of LiFeAs 
is very different from that of BaFe$_2$(As,P)$_2$.

\begin{figure}[!htb]
\includegraphics[width=0.9\linewidth]{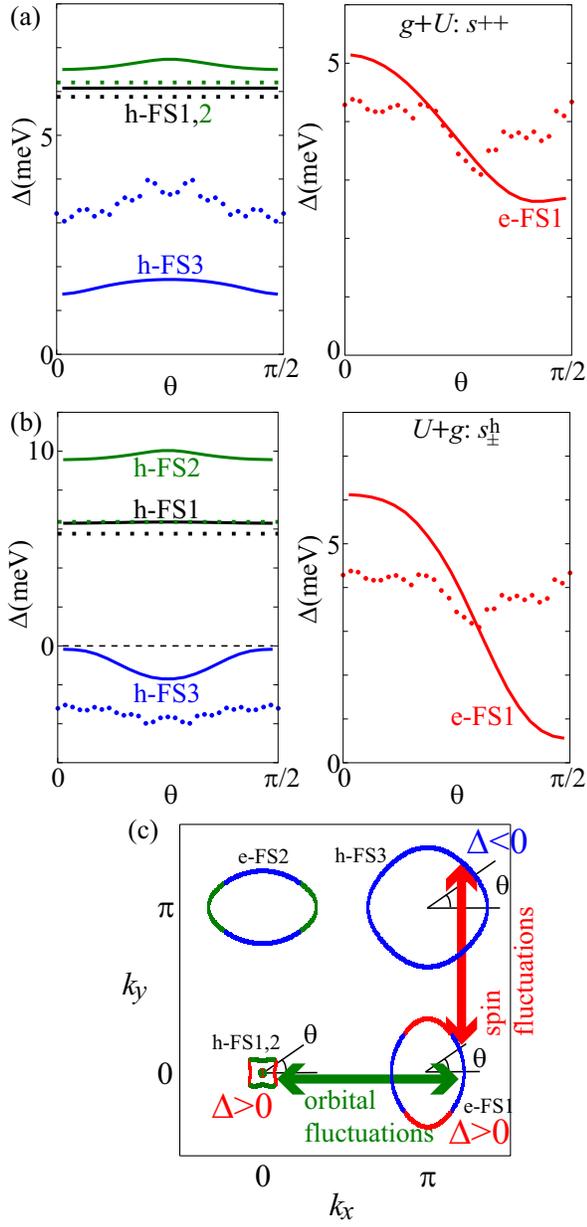}
\caption{
(Color online)
Obtained gap functions in the case of $U\ne0$ and $g\ne0$:
(a) The $s_{++}$-wave state for $U=0.200$ and $g=0.125$ 
in the $k_z=\pi$-plane. 
The eigenvalue is $\lambda_E=0.47$.
(b) The hole-$s_\pm$-wave state for $U=0.380$ and $g=0.122$ 
in the $k_z=\pi$-plane, in which only the gap function on the h-FS3 is negative.
The eigenvalue is $\lambda_E=0.20$.
(c) Origin of the hole-$s_\pm$-wave state due to the coexistence of 
the ``orbital-fluctuations among $(d_{xz},d_{yz})$- and $d_{xy}$-orbitals''
and the ``spin-fluctuations on the $d_{xy}$-orbital''.
}
\label{fig:gap3}
\end{figure}

Figure \ref{fig:gap3} (a) shows the 
obtained gap functions in the case of $g=0.125$ and $U=0.200$.
The obtained Stoner factors are $\a_c=0.98$ and $\a_s=0.45$.
In this case, the orbital fluctuations are much larger than 
the spin fluctuations, and therefore we obtain the $s_{++}$-wave state.
Except for h-FS3, the obtained gap structures are similar to those of 
the ``pure $s_{++}$-wave state'' without $U$ in Fig. \ref{fig:gap1}.
Due to the moderate spin fluctuations on the $d_{xy}$-orbital,
the anisotropy of $\Delta^{\rm h}_3$ is enlarged, 
consistently with experimental results.

If we increase the value of $U$ further, 
we obtain a highly nontrivial gap structure with sign-reversal
within the h-FSs:
Figure \ref{fig:gap3} (b) shows the 
obtained gap functions in the case of $g=0.122$ and $U=0.380$
($\a_c=0.98$ and $\a_s=0.85$).
Here, only $\Delta^{\rm h}_3$ is negative.
In this ``hole-$s_\pm$-wave state'' with 
``sign-reversal within hole-pockets'',
the obtained gap structures of $\Delta^{\rm h}_{1,2}$ and $\Delta^{\rm e}_{1,2}$
are qualitatively similar to those in the $s_{++}$-wave state in
Fig. \ref{fig:gap1}.
On the other hand, $\Delta^{\rm h}_3$ becomes very anisotropic, 
similarly to $\Delta^{\rm h}_3$ in the $s_\pm$-wave state in Fig. \ref{fig:gap2}.

We discuss the reason why hole-$s_\pm$-wave is realized 
by the coexistence of orbital and spin fluctuations:
In the present hole-$s_\pm$-wave state,
as shown in Fig.  \ref{fig:gap3} (c),
$\Delta^{\rm h}_{1,2} \cdot \Delta^{\rm e}_{1,2}$ is positive 
due to orbital fluctuations,
whereas $\Delta^{\rm h}_3 \cdot \Delta^{\rm e}_{1,2}$ is negative
due to spin fluctuations.
The obtained gap structure is qualitatively consistent with 
ARPES measurement in Ref. \cite{Borisenko-LiFeAs},
although the gap structures of the $s_{++}$-wave state in 
Fig. \ref{fig:gap1} are more consistent with experiments.
The present mechanism of the 
``sign-reversal within hole-pockets'' due to orbital+spin fluctuations 
would be realized in other Fe-based superconductors.
In fact, the hole-$s_\pm$-wave state 
was first discussed in Ba$_{1-x}$K$_x$Fe$_2$As$_2$
based on the thermal conductivity and penetration depth measurements
\cite{Watanabe-shpm},
in addition to the recent ARPES study \cite{Ding-shpm}.

Finally, we discuss on other theoretical works
which predict the sign-reversal within hole-pockets.
The hole-$s_\pm$-wave state was first discussed
by the authors in Ref. \cite{Chubukov-spmh},
assuming the repulsive interaction between h-FSs and e-FSs
in addition to the repulsive pairing interaction within the h-FSs.
For LiFeAs, similar scenario was discussed in Ref. \cite{Ahn},
by introducing competing repulsive interactions,
although the repulsive interaction within the h-FSs 
is much weaker within the RPA because of the ill-nesting.
Also, the authors in Ref. \cite{Yin} discussed the 
orbital antiphase $s^{+-}$ state, in which 
the sign-reversal within hole-pocket is realized due to 
the strong repulsion between $d_{xy}$ and $d_{xz,yz}$ orbitals.
In this state, the gap on e-FS is nodal in the unfolding picture,
whereas it is fully-gapped in the present hole-$s_\pm$ state
in Fig. \ref{fig:gap3} (b).

References \cite{Chubukov-spmh,Ahn,Yin} considered the 
competition between two kinds of repulsive interactions.
In contrast, in the present paper, the  hole-$s_\pm$-wave state
is explained in terms of the cooperation between the 
``attractive interaction among $(d_{xz},d_{yz})$- and $d_{xy}$-orbitals''
and ``repulsive interaction on the $d_{xy}$-orbital''.

\section{Self-Consistent VC+$\Sigma$ (SC-VC$_\Sigma$) Method
for the Hubbard model ($g=0$)}
\label{sec:SCVC}

In previous sections, we studied the 
extended Hubbard model with multiorbital Coulomb interaction
($U$, $U'$, $J=(U-U')/2$) and quadrupole interaction ($g$).
Here, orbital (spin) fluctuations are induced by $g$ ($U$)
and inter-orbital (intra-orbital) nesting of the FSs.
Orbital fluctuations are the driving force of the 
fully-gapped $s_{++}$-wave state, and the coexistence of
orbital and spin fluctuations gives rise to the hole-$s_\pm$ state
with the sign-reversal within the hole-pockets.

In Ref. \cite{Kontani-RPA},
we had shown that $g$ is induced by in-plane Fe-ion oscillations.
Consistently, kink structure in the quasiparticle dispersion
due to Fe-ion oscillations is observed experimentally in LiFeAs 
\cite{Borisenko-LiFeAs}.
Later, we found that $g$ is also induced by the 
Coulomb interaction (without $e$-ph interaction) beyond the RPA:
It was revealed that the Aslamazov-Larkin type vertex correction (AL-VC)
produce large effective quadrupole interaction $g$ \cite{Onari-SCVC}.
By solving the model for LaFeAsO
using the self-consistent vertex correction (SC-VC) method,
we obtain strong developments of 
$\chi_{x^2-y^2}^Q(\q)$ at $q=(0,0)$ and 
$\chi_{xz,yz}^Q(\q)$ at $q=(0,\pi),(\pi,0)$
\cite{Onari-SCVC}.
The former fluctuations explain the orthorhombic structure 
transition in mother compounds
\cite{Kontani-Raman}.

\begin{figure}[!htb]
\includegraphics[width=0.99\linewidth]{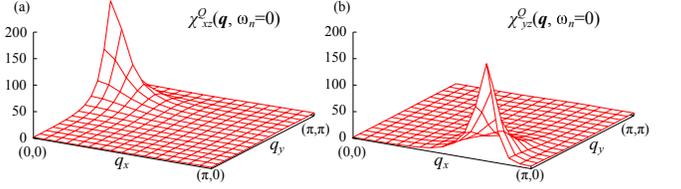}
\caption{
(Color online)
The quadruple susceptibility $\chi_{xz}^Q(\q)$ and $\chi_{yz}^Q(\q)$
obtained by the SC-VC$_\Sigma$ method based on the Hubbard model.
Other quadruple fluctuations are small.
The used parameters are $U=0.96$, $J/U=0.13$, and $g=0$.
The Stoner factors are $\a_c=0.97$ and $\a_s=0.87$.
}
\label{fig:SCVC}
\end{figure}

In this section, we analyze the tight-binding Hubbard model ($g=0$) 
for LiFeAs using the self-consistent VC+$\Sigma$ (SC-VC$_\Sigma$) method,
which was used in Refs. \cite{Onari-SCVCS,Onari-Hdoped}.
To simply the numerical calculation, we study the $k_z=\pi$-plane
of the present three-dimensional LiFeAs model.
In the SC-VC$_\Sigma$ method,
the self-energy matrix $\hat{\Sigma}$ is given by the 
the one-loop approximation:
\begin{eqnarray}
\Sigma_{lm}(k)=T\sum_{q}\sum_{l',m'}V^\Sigma_{ll',mm'}(q)G_{l'm'}(k-q),
\label{eqn:self}
\end{eqnarray}
where $G$ is the 
full Green function with self-energy given as
$\{\hat G(k)\}^{-1} = \{{\hat G}^{(0)}(k)\}^{-1}-{\hat \Sigma}(k)$.
$\hat{V}^\Sigma(q)$ is the effective interaction for the self-energy:
$\displaystyle
\hat{V}^\Sigma(q)=\frac{3}{2}\hat{\Gamma}^s\hat{\chi}^s(q)\hat{\Gamma}^s
+\frac{1}{2}\hat{\Gamma}^c\hat{\chi}^c(q)\hat{\Gamma}^c
-\frac{1}{4}(\hat{\Gamma}^c-\hat{\Gamma}^s)\hat{\chi}^0(q)(\hat{\Gamma}^c-\hat{\Gamma}^s)-\frac{1}{8}(\hat{\Gamma}^c+\hat{\Gamma}^s)\hat{\chi}^0(q)(\hat{\Gamma}^c+\hat{\Gamma}^s)$.
The third and fourth terms of the right hand side in $\hat{V}^\Sigma(q)$ are
required to cancel the double counting in the 2nd order diagrams.

The susceptibility for the charge (spin) sector is
\begin{eqnarray} 
&&\hat{\chi}^{\mathrm{c(s)}} \left( q \right) = \frac{\hat{\Phi}^{c(s)} 
\left( q \right)}{\hat{1} - \hat{\Gamma}^{\mathrm{c(s)}} 
\hat{\Phi}^{c(s)} \left( q \right)},
\label{eqn:chic-VC}
\end{eqnarray}
where ${\hat \Phi}^{c(s)}(q)\equiv {\hat \chi}_\Sigma^{(0)}(q)+{\hat X}^{c(s)}(q)$
is the irreducible susceptibility.
${\hat \chi}_\Sigma^{(0)}(q)$ is given by eq. (\ref{eqn:chi0}),
by replacing ${\hat G}^{(0)}$ with the full Green function ${\hat G}$.
${\hat X}^{c(s)}(q)$ is the VC for the charge (spin) sector.
In the SC-VC$_\Sigma$ method, 
we calculate the VC up to the second-order terms with respect to 
the susceptibility $\chi^{s,c}$.
The second-order term (=Aslamazov-Larkin term) is always dominant over the 
first-order term (=Maki-Thompson term), and the 
AL-VC for the charge sector is given as \cite{Onari-SCVC}
\begin{eqnarray}
&&X_{ll',mm'}^{c}(q)
\nonumber \\
&&\ \ \ \ =\frac{T}2\sum_{k}\sum_{a\sim h}
\Lambda_{ll',ab,ef}(q;k)\{  {V}_{ab,cd}^c(k+q){V}_{ef,gh}^c(-k)
\nonumber \\
& &\ \ \ \ \ \ +3{V}_{ab,cd}^s(k+q){V}_{ef,gh}^s(-k) \}
\Lambda_{mm',cd,gh}'(q;k) ,
 \label{eqn:ALexample}
\end{eqnarray}
where 
${\hat V}^{s,c}(q)\equiv{\hat \Gamma}^{s,c}
+ {\hat\Gamma}^{s,c}{\hat\chi}^{s,c}(q){\hat\Gamma}^{s,c} $, 
${\hat \Lambda}(q;k)$ and ${\hat \Lambda}'(q;k)$ are the three-point vertex
made of three Green functions given in Ref. \cite{Onari-SCVC}.
We include all $U^2$-terms without the double counting
to obtain reliable results. 
Here, we neglect $\hat{X}^{{\rm AL},s}$ because the
contribution of $\hat{X}^{{\rm AL},s}$ is much smaller than that of 
$\hat{X}^{{\rm AL},c}$ \cite{Onari-SCVC,Ohno-SCVC,Tsuchiizu}.
Also, we use $\hat{G}^{(0)}$ 
in calculating ${\hat \Lambda}$ and ${\hat \Lambda}'$
since they are underestimated at high temperatures ($T\gg0.01$)
due to large quasiparticle damping
Im$\Sigma(\bm{q},-i\delta) \propto T$.

In the SC-VC$_\Sigma$ method, we solve 
eqs. (\ref{eqn:self})-(\ref{eqn:ALexample}) self-consistently.
Here, we study the two-dimensional model given by 
the $k_z=\pi$ plane of LiFeAs using the SC-VC$_\Sigma$ method.
Figure \ref{fig:SCVC} shows the obtained
quadruple susceptibility $\chi_{xz}^Q(\q)$ and $\chi_{yz}^Q(\q)$.
The used parameters are $U=0.96$, $J/U=0.13$, $g=0$, and $T=0.02$.
The obtained $\chi_{xz,yz}^Q(\q)$ shows incommensurate peak structure,
reflecting the bad nesting of the FSs in LiFeAs \cite{comment}.
In highly contrast to the case of LaFeAsO \cite{Onari-SCVC},
$\chi_{x^2-y^2}^Q({\bf 0})$ in the present model is very small,
consistently with the absence of structure transition in LiFeAs.
(We verified that very similar result is obtained by 
the SC-VC method (without self-energy correction)
by putting $J/U\lesssim0.09$.)
Thus, the quadrupole interaction in eq. (\ref{eqn:Hint}) 
is derived from the VC due to Coulomb interaction 
in addition to the e-ph interaction.

Theoretically, the $O_{xz/yz}$ type quadrupole fluctuations 
are easily realized because of the good inter-orbital nesting of the FSs. 
They are produced by taking account of the small quadrupole interaction $g$ 
and/or the AL term due to Coulomb interaction. 
In fact, both AL term and $g$ contribute to the $O_{xz/yz}$-type 
quadrupole fluctuations cooperatively \cite{Onari-SCVC}, 
indicating that the phenomenological interaction $g$ 
can be used as a substitute for the AL term.

In Ref. \cite{Onari-Hdoped},
we solved the gap equation based on the ``two-dimensional'' model for
LaFeAsO$_{1-x}$H$_x$ using the SC-VC$_\Sigma$ method,
and obtained various types of $s$-wave superconducting states,
like $s_{++}$-, $s_\pm$-, and hole-$s_\pm$-wave states,
due to the cooperation of orbital and spin fluctuations.
It is our future problem to study the 
``three-dimensional'' gap structure of LiFeAs
based on the SC-VC$_\Sigma$ method.
In LiFeAs, it is naively expected that 
strong incommensurate orbital fluctuations shown in Fig. \ref{fig:SCVC} 
produces large $\Delta_{1,2}^{\rm h}$ like in Fig. \ref{fig:gap1} (a),
since h-FS1,2 (made of $d_{xz,yz}$-orbitals) and 
e-FS1,2 (made of $d_{xy}$-orbial) are connected by 
these incommensurate orbital fluctuations.

\section{Summary}
\label{sec:discussion}

In this paper,
we studied the three-dimensional five-orbital model of LiFeAs
based on the recently-developed orbital-spin fluctuation theories
\cite{Kontani-RPA,Onari-SCVC}.
It is found that the experimentally observed gap structure of LiFeAs
in Ref. \cite{Borisenko-LiFeAs}
is quantitatively reproduced in terms of the orbital-fluctuation mechanism.
Especially, the largest gap on h-FS1 and h-FS2 in Fig. \ref{fig:FS} (b)
is naturally reproduced by the inter-orbital fluctuations,
as demonstrated in Figs. \ref{fig:gap1} (a) and \ref{fig:gap1-90},
whereas it is unable to be explained by the spin fluctuation scenario.
Therefore, the largest gap on h-FS1,2 
is the hallmark of the orbital-fluctuation-mediated superconductivity 
in LiFeAs.
Also, the orbital-independent isotropic gap (absence of horizontal node) 
on h-FSs in Ba122 \cite{Shimo-Science,Yoshida} and Sr122 
indicates the important role of orbital fluctuations
on the pairing mechanism \cite{Saito-loop}.

\begin{figure}[!htb]
\includegraphics[width=0.99\linewidth]{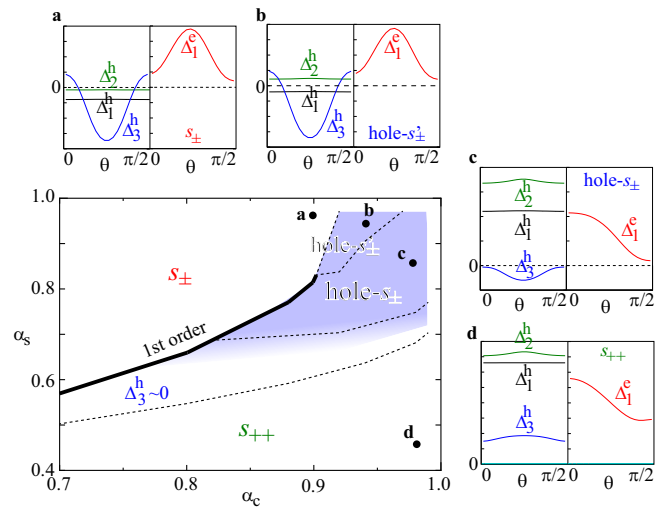}
\caption{
(Color online)
$\a_c$-$\a_s$ phase diagram of the gap structure in LiFeAs.
The gap structure at each point {\bf a}$\sim${\bf d} 
is shown in the figure.
Each $s_\pm$-wave, $s_{++}$-wave, and hole-$s_\pm$-wave state
is realized in wide parameter region.
In the region ``$\Delta_3^{\rm h}\sim0$, 
the gaps on other FSs have the same sign, so 
nearly $s_{++}$-wave state is realized.
In the ``hole-$s_\pm'$-wave gap'' state at point {\bf b},
$\Delta_1^{\rm h} \cdot \Delta_2^{\rm h}$ is negative, and
both $|\Delta_1^{\rm h}|$ and $|\Delta_2^{\rm h}|$ are very small.
}
\label{fig:phase}
\end{figure}

When orbital and spin fluctuations coexist,
the ``hole-$s_\pm$-wave state'' is obtained, in which 
only the gap of the largest $d_{xy}$-orbital hole-pocket
is sign-reversed.
We expect that the present mechanism of the 
``sign-reversal within hole-pockets'' due to orbital+spin fluctuations 
would be realized in other Fe-based superconductors,
although LiFeAs might not be the case.
In fact, the realization of the hole-$s_\pm$-wave state 
was first discussed in Ba$_{1-x}$K$_x$Fe$_2$As$_2$
based on the thermal conductivity and penetration depth measurements
\cite{Watanabe-shpm}.
The hole-$s_\pm$-wave is naturally realized under the coexistence of
the ``spin-fluctuations on the $d_{xy}$-orbital'' and 
the ``orbital-fluctuations among the $(d_{xz},d_{yz})$- and $d_{xy}$-orbitals''.

Figure \ref{fig:phase} shows the obtained 
$\a_c$-$\a_s$ phase diagram of the gap structure in LiFeAs.
As expected, the $s_\pm$-wave state ($s_{++}$-wave state) is realized 
for wide ragion of $\a_s>\a_c$ ($\a_c>\a_s$).
When both $\a_s$ and $\a_c$ are close to unity,
we obtain the hole-$s_\pm$-wave gap in a wide region.
The gap structure at each point {\bf a}$\sim${\bf d}
is shown in the figure.
In the region ``$\Delta_3^{\rm h}\sim0$'', 
obtained $\Delta_3^{\rm h}(\theta)$ is nodal and very small in magnitude,
and it is close to the $s_{++}$-wave state 
in that other gaps are positive and large.
In the ``hole-$s_\pm'$-state'' at point {\bf b},
$\Delta_1^{\rm h}$ and $\Delta_2^{\rm h}$ are opposite in sign, and
both $|\Delta_1^{\rm h}|$ and $|\Delta_2^{\rm h}|$ are very small.
Therefore, various types of $s$-wave gap structure are realized 
due to the cooperation of orbital and spin fluctuations.

We also applied the SC-VC$_\Sigma$ method to the 
Hubbard model of LiFeAs, and obtained the 
strong development of antiferro-orbital fluctuations
due to the AL-type VC.
In contrast, the ferro-orbital fluctuations remain small
contrary to the previous study for La1111 \cite{Onari-SCVC},
consistently with the absence of orthorhombic
structure transition in LiFeAs.
It is our important future issue to study the 
superconducting state of LiFeAs based on the SC-VC$_\Sigma$ method.

\acknowledgements
This study has been supported by Grants-in-Aid for Scientific 
Research from MEXT of Japan.
S.V.B. and V.B.Z. acknowledge support under Grants No. ZA 654/1-1, No. BO1912/2-2, and No. BE1749/13.
Part of numerical calculations were
performed on the Yukawa Institute Computer Facility.

\appendix
\section{Gap Structure due to Moderate Orbital and Spin Fluctuations}
\label{sec:Appendix-GAP}

In Sec. \ref{sec:RPA},
we have shown the gap structures of the 
$s_{++}$-wave and $s_\pm$-wave states in the presence of 
very large orbital and spin fluctuations; $\a_{\rm c,s}=0.98$.
However, we have very little experimental information 
on the strength of fluctuations in LiFeAs.
In fact, the spin fluctuations are moderate
according to NMR measurement \cite{NMR-LiFeAs}
and neutron scattering measurement 
\cite{neutron-LiFeAs1,neutron-LiFeAs2,neutron-LiFeAs3}.
In this Appendix, we analyze the gap equation 
for smaller orbital and spin fluctuations,
and show that the obtained gap structures in Sec. \ref{sec:RPA}
are essentially unchanged even when the fluctuations are moderate.

\begin{figure}[!htb]
\includegraphics[width=0.9\linewidth]{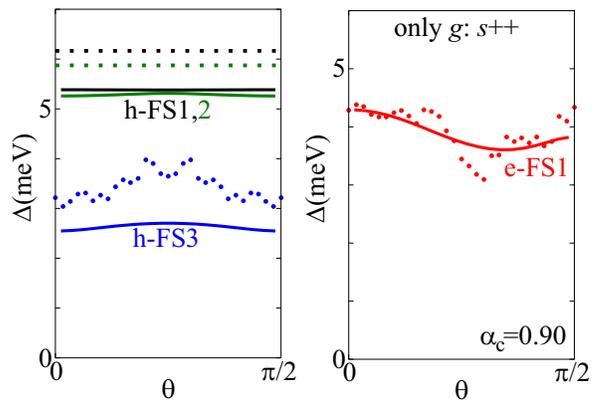}
\caption{
(Color online)
Obtained $s_{++}$-wave gap structure
for $U=0$ and $g=0.118$ ($\a_{\rm c}=0.90$) in the $k_z=\pi$-plane. 
The eigenvalue is $\lambda_E=0.34$.
The dotted lines represent the experimental data
given by the ARPES measurement in Ref. \cite{Borisenko-LiFeAs}.
}
\label{fig:gap1-90}
\end{figure}

Figure \ref{fig:gap1-90} shows the 
$s_{++}$-wave gap functions for $g=0.118$ and $U=0$.
In this case, $\a_{\rm c}=0.90$ and ${\rm max}_\q\chi^Q_{xz}(\q,0)\approx38$.
The obtained gap structure is very similar to that in 
Fig. \ref{fig:gap1} (a) for  $\a_{\rm c}=0.98$.
Especially, experimentally observed local maximum at $\theta=\pi/2$
on the e-FS3 is well reproduced in Fig. \ref{fig:gap1-90}.
Thus, the $s_{++}$-wave gap structure is essentially unchanged 
for $\a_{\rm c}\ge0.90$,
although the eigenvalue $\lambda_E$ increases as $\a_{\rm c}$
approaches unity.
The obtained eigenvalue $\lambda_E=0.34$ is relatively large,
which means that moderate orbital fluctuations ($\a_{\rm c}=0.90$)
would be enough to induce the superconductivity.

\begin{figure}[!htb]
\includegraphics[width=0.9\linewidth]{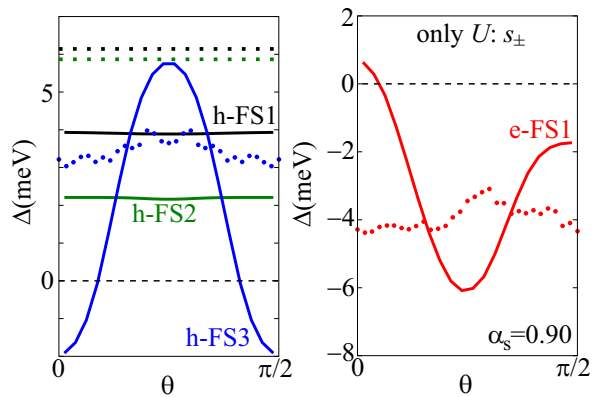}
\caption{
(Color online)
Obtained $s_\pm$-wave gap structure
for $U=0.403$ and $g=0$ ($\a_{\rm s}=0.90$) in the $k_z=\pi$-plane.
The eigenvalue is $\lambda_E=0.12$.
}
\label{fig:gap2-90}
\end{figure}

Figure \ref{fig:gap2-90} shows the 
$s_\pm$-wave gap functions for $U=0.403$ and $g=0$.
In this case, $\a_{\rm s}=0.90$ and ${\rm max}_\q\chi^s(\q,0)\approx35$.
As for the hole-pockets,
the obtained gap functions are essentially similar to those
in Fig. \ref{fig:gap2} for $\a_{\rm s}=0.98$,
except that $\Delta^{\rm h}_{1,2}$ becomes relatively large.
As for the electron-pocket, the nodal gap appears on the e-FSs, 
although it is inconsistent with experiments.
Thus, the overall experimental data is difficult to be explained 
by the spin fluctuation theory for $\a_{\rm s}\ge0.90$.

\section{Five-Orbital Tight-Binding Model for LiFeAs}
\label{sec:Appendix-TB}

{\scriptsize
\begin{table*}
\begin{tabular}{c|rrrrrrrrrccc}
\backslashbox{\small($l$,$m$)}{\small $\bm{R}$}
        & [0,0,0]& [1,0,0]& [1,1,0]& [2,0,0]& [2,1,0]& [2,2,0]& [0,0,1]& [1,0,1]& [2,0,1]&$\sigma_y$&$I$&$\sigma_d$\\ 
\hline
  (1,1) &$-$0.305&        &        &        &        &        &        &        &        & + & + &  + \\
  (1,2) &        &        &$-$0.101&        &        &        &        &        &        &$-$&$-$&  +(1,3) \\
  (1,3) &        &   0.100&$-$0.101&        &        &        &        &        &        & + &$-$&  +(1,2) \\
  (1,4) &        &        &$-$0.090&        &        &        &        &        &        &$-$& + &  + \\
  (1,5) &        &$-$0.162&        &        &        &        &        &        &        & + & + &$-$ \\
  (2,2) &$-$0.008&$-$0.050&   0.152&$-$0.004&$-$0.040&$-$0.005&$-$0.003&$-$0.012&        & + & + &  +(3,3) \\
  (2,3) &        &        &   0.090&        &        &        &        &        &        &$-$& + &  + \\
  (2,4) &        &$-$0.155&$-$0.064&        &        &        &        &        &        & + &$-$&  +(3,4) \\
  (2,5) &        &        &$-$0.010&        &        &        &        &        &        &$-$&$-$&$-$(3,5) \\
  (3,3) &$-$0.008&$-$0.210&   0.152&$-$0.051&   0.053&$-$0.005&$-$0.003&        &   0.011& + & + &  +(2,2) \\
  (3,4) &        &        &$-$0.064&        &        &        &        &        &        &$-$&$-$&  +(2,4) \\
  (3,5) &        &   0.193&   0.010&        &        &        &        &        &        & + &$-$&$-$(2,5) \\
  (4,4) &   0.020&   0.019&   0.030&$-$0.010&$-$0.004&        &   0.011&   0.004&        & + & + &  + \\
  (4,5) &        &        &        &        &        &        &        &        &        &$-$& + &$-$ \\
  (5,5) &$-$0.261&   0.223&   0.070&        &        &        &        &        &        & + & + &  + \\
\end{tabular}
\caption{
Hopping integrals for $\bm{R}=(x, y, z)$
for the present five-orbital model for LiFeAs.
Notations are the same as those introduced in 
Refs. \cite{Kuroki,Miyake}.
$\sigma_y$, $I$, and 
$\sigma_d$ corresponds to $t(x, -y, z;l,m)$, 
$t(-x, -y, z;l,m)$, and $t(y, x, z;l,m)$, respectively.
Here, `$\pm$' and `$\pm (l',m')$' in the row of 
$(l,m)$ mean that the corresponding  
hopping is equal to $\pm t(x, y, z; l, m)$ and  
$\pm t(x, y, z; l', m')$, respectively. 
Notice also $t(\bm{R}; l, m) = t(-\bm{R}; m, l)$.
\label{tab1}
}
\end{table*}
%
}


Here, we explain the five-orbital tight-binding model for LiFeAs,
which is given in unfolding the ten-orbital model
given in Ref. \cite{Hirschfeld-LiFeAs}.
The ten-orbital model in Ref. \cite{Hirschfeld-LiFeAs} 
is obtained by fitting the experimental 
band structure of LiFeAs in Ref. \cite{Borisenko-LiFeAs}
near the Fermi level.
The five-orbital model (single Fe unit cell) is obtained by 
unfolding the ten-orbital model (two-Fe unit cell),
according to the procedure in Ref. \cite{Miyake}.
The FSs of both models are shown in Fig. \ref{fig:FS} (a) and (b), respectively.
Both models are equivalent, and the former model is 
more convenient for the numerical study.
The FSs and band structrue are given in Fig. \ref{fig:FS}.
This experimental FSs of LiFeAs are very different from the FSs 
given by the density functional theory (DFT), 
in which FS1,2 predicted by the DFT are much larger.
Better agreement between theory and ARPES is achieved 
by the LDA+DMFT study \cite{DMFT}, since the FS1,2
shrinks due to the orbital-dependent self-energy that is absent in the LDA.
As for the de Haas-van Alphen (dHvA) measurements,
Ref. \cite{dHvA1} showed reasonable agreement with the DFT for the e-FSs,
and Ref. \cite{dHvA2} reported the presence of very small
three-dimensional hole-pockets, which would corresponds to 
h-FS1,2 in Fig. \ref{fig:FS}.
The hopping parameters of the present model,
$t^{l,m}(\bm{R}_{a})$ in eq. (\ref{eqn:H0}), are listed in Table I.

Based on the LDA bandstructure,
$\Delta^{\rm h}_{1,2}$ in the spin fluctuation mediated $s_\pm$-wave state
can become as large as other gaps,
as discussed in Refs. \cite{Hirschfeld-LiFeAs,Platt}.
However, $\Delta^{\rm h}_{1,2}$ in the $s_\pm$-wave state
becomes very small based on the ``experimental bandstructure'',
as shown in Ref. \cite{Hirschfeld-LiFeAs} and in the present paper.





\begin{thebibliography}{99}

\bibitem{NMR-Fujiwara}
N. Fujiwara, S. Tsutsumi, S. Iimura, S. Matsuishi, H. Hosono, Y. Yamakawa, and H. Kontani, 
Phys. Rev. Lett. {\bf 111}, 097002 (2013).

\bibitem{NMR-Ishida}
Y. Nakai, T. Iye, S. Kitagawa, K. Ishida, H. Ikeda, S. Kasahara, H. Shishido, T. Shibauchi, Y. Matsuda, and T. Terashima,
 Phys. Rev. Lett. {\bf 105}, 107003 (2010).

\bibitem{NMR-Zheng}
R. Zhou, Z. Li, J. Yang, D. L. Sun, C. T. Lin, and G. Zheng, Nat. Commun {\bf 4}, 3265, (2013).

\bibitem{C66-Fer}
R.M. Fernandes, L. H. VanBebber, S. Bhattacharya, P. Chandra, 
V. Keppens, D. Mandrus, M.A. McGuire, B.C. Sales, A.S. Sefat 
and J. Schmalian, 
Phys. Rev. Lett. {\bf 105}, 157003 (2010). 

\bibitem{C66-Yoshizawa}
M. Yoshizawa, R. Kamiya, R. Onodera, Y. Nakanishi, K. Kihou, H. Eisaki 
and C. H. Lee, 
Phys. Soc. Jpn. {\bf 81}, 024604 (2012).

\bibitem{C66-Bohmer}
A. E. B\"{o}hmer, P. Burger, F. Hardy, T. Wolf, P. Schweiss, R. Fromknecht, M. Reinecker, W. Schranz, and C. Meingast, 
Phys. Rev. Lett. {\bf 112}, 047001 (2014). 

\bibitem{Raman}
Y. Gallais, R. M. Fernandes, I. Paul, L. Chauviere, Y.-X. Yang, 
M.-A. Measson, M. Cazayous, A. Sacuto, D. Colson, and A. Forget,
Phys. Rev. Lett. {\bf 111}, 267001 (2013).

\bibitem{FeSe1}
Y. Mizuguchi and Y. Takano, J. Phys. Soc. Jpn. {\bf 79}, 102001 (2010).

\bibitem{Borisenko-various}
A. A. Kordyuk, V. B. Zabolotnyy, D. V. Evtushinsky, A. N. Yaresko, B. B\"{u}chner,  and S. V. Borisenko, J. Supercond. Nov. Magn. {\bf 26}, 2837 (2013).

\bibitem{Borisenko-FeSe}
J. Maletz, V. B. Zabolotnyy, D. V. Evtushinsky, S. Thirupathaiah, A. U. B. Wolter, L. Harnagea, A. N. Yaresko, A. N. Vasiliev, D. A. Chareev, E. D. L. Rienks, B. B\"{u}chner, and S. V. Borisenko,  
Phys. Rev. B {\bf 89}, 220506(R) (2014).

\bibitem{Kuroki}
K. Kuroki, S. Onari, R. Arita, H. Usui, Y. Tanaka, H. Kontani, and H. Aoki, 
Phys. Rev. Lett. {\bf 101}, 087004 (2008).

\bibitem{Mazin}
I. I. Mazin, D.J. Singh, M.D. Johannes, and M.H. Du,
Phys. Rev. Lett. {\bf 101} (2008) 057003.

\bibitem{Hirschfeld}
P. J. Hirschfeld, M. M. Korshunov, and I. I. Mazin, 
Rep. Prog. Phys. {\bf 74}, 124508 (2011).

\bibitem{Chubukov}
A. V. Chubukov, D. V. Efremov, and I. Eremin, 
Phys. Rev. B {\bf 78}, 134512 (2008).

\bibitem{Kontani-RPA}
H. Kontani and S. Onari, 
Phys. Rev. Lett. {\bf 104}, 157001 (2010).

\bibitem{Onari-SCVC}
S. Onari and H. Kontani, 
Phys. Rev. Lett. {\bf 109}, 137001 (2012).

\bibitem{Sato-imp}
M. Sato, Y. Kobayashi, S. C. Lee, H. Takahashi, E. Satomi, and Y. Miura,
J. Phys. Soc. Jpn. {\bf 79} (2009) 014710;
S. C. Lee, E. Satomi, Y. Kobayashi, and M. Sato,
J. Phys. Soc. Jpn. {\bf 79} (2010) 023702.

\bibitem{Li-imp}
J. Li, Y.F. Guo, S.B. Zhang, J. Yuan, Y. Tsujimoto, X. Wang, C.I. Sathish, 
Y. Sun, S. Yu, W. Yi, K. Yamaura, E. Takayama-Muromachi, Y. Shirako, 
M. Akaogi, and H. Kontani, Phys. Rev. B {\bf 85}, 214509 (2012).

\bibitem{Nakajima-imp}
Y. Nakajima, T. Taen, Y. Tsuchiya, T. Tamegai, H. Kitamura, and T. Murakami,
Phys. Rev. B {\bf 82}, 220504 (2010).

\bibitem{Onari-imp}
S. Onari and H. Kontani, 
Phys. Rev. Lett. {\bf 103}, 177001 (2009).

\bibitem{Yamakawa-imp}
Y. Yamakawa, S. Onari and H. Kontani, 
Phys. Rev. B {\bf 87}, 195121 (2013).

\bibitem{Onari-neutron}
S. Onari, H. Kontani, and M. Sato,
Phys. Rev. B {\bf 81}, 060504(R) (2010);
S. Onari and H. Kontani, Phys. Rev. B {\bf 84}, 144518 (2011)

\bibitem{Borisenko-LiFeAs}
S.V. Borisenko, V.B. Zabolotnyy, A.A. Kordyuk, D.V. Evtushinsky, 
T.K. Kim, I.V. Morozov, R. Follath and B.B\"{u}chner,
Symmetry {\bf 4}, 251 (2012).

\bibitem{Ding-LiFeAs}
K. Umezawa, Y. Li, H. Miao, K. Nakayama, Z.-H. Liu, P. Richard, T. Sato, J. B. He, D.-M. Wang, G. F. Chen, H. Ding, T. Takahashi, and S.-C. Wang,
Phys. Rev. Lett. {\bf 108}, 037002 (2012).

\bibitem{Miyake}
T. Miyake, K. Nakamura, R. Arita and M. Imada, 
J. Phys. Soc. Jpn. {\bf 79}, 044705 (2010).

\bibitem{NMR-LiFeAs}
Li Zheng, Ooe Yosuke, Wang Xian-Cheng, Liu Qing-Qing, Jin Chang-Qing, Ichioka Masanori, Zheng Guo-qing,  J. Phys. Soc. Jpn {\bf 79}, 083702 (2010).

\bibitem{neutron-LiFeAs1}
N. Qureshi, P. Steffens, Y. Drees, A.C. Komarek, D.
Lamago, Y. Sidis, L. Harnagea, H.-J. Grafe, S. Wurmehl,
B. B{\"u}chner, and M. Braden, Phys. Rev. Lett. {\bf 108}, 117001 (2012).

\bibitem{neutron-LiFeAs2}
J. Knolle, V. B. Zabolotnyy, I. Eremin, S. V. Borisenko, N. Qureshi, M. Braden, D. V. Evtushinsky, T. K. Kim, A. A. Kordyuk, S. Sykora, Ch. Hess, I. V. Morozov, S. Wurmehl, R. Moessner, and B. B\"{u}chner,
Phys. Rev. B {\bf 86}, 174519 (2012).

\bibitem{neutron-LiFeAs3}
A. Taylor, et al., Phys. Rev. B {\bf 83}, 220514 (2011).

\bibitem{Hirschfeld-LiFeAs}
Y. Wang, A. Kreisel, V. B. Zabolotnyy, S. V. Borisenko, B. B\"{u}chner, T. A. Maier, P. J. Hirschfeld, and D. J. Scalapino,
Phys. Rev. B {\bf 88}, 174516 (2013).

\bibitem{QPI}
M. P. Allan, A. W. Rost, A. P. Mackenzie, Yang Xie, J. C. Davis, K. Kihou,
 C. H. Lee, A. Iyo, H. Eisaki, and T.-M. Chuang,
Science {\bf 336}, 563 (2012).

\bibitem{Watanabe-shpm}
D. Watanabe, T. Yamashita, Y. Kawamoto, S. Kurata, Y. Mizukami, T. Ohta, S. Kasahara, M. Yamashita, T. Saito, H. Fukazawa, Y. Kohori, S. Ishida, K. Kihou, C. H. Lee, A. Iyo, H. Eisaki, A. B. Vorontsov, T. Shibauchi, and Y. Matsuda,
Phys. Rev. B {\bf 89}, 115112 (2014).

\bibitem{Ding-shpm}
P. Zhang, P. Richard, T. Qian, X. Shi, J. Ma, L.-K. Zeng, X.-P. Wang, E. Rienks, C.-L. Zhang, Pengcheng Dai, Y.-Z. You, Z.-Y. Weng, X.-X. Wu, J. P. Hu, and H. Ding,  arXiv:1312.7064.

\bibitem{Suzuki}
K. Suzuki, H. Usui, and K. Kuroki,
J. Phys. Soc. Jpn. {\bf 80}, 013710 (2011).

\bibitem{Saito-loop}
T. Saito, S. Onari, and  H. Kontani,
Phys. Rev. B {\bf 88}, 045115 (2013).

\bibitem{Shimo-Science}
T. Shimojima, F. Sakaguchi, K. Ishizaka, Y. Ishida, T. Kiss, M. Okawa, 
T. Togashi, C.-T. Chen, S. Watanabe, M. Arita, K. Shimada, H. Namatame, 
M. Taniguchi, K. Ohgushi, S. Kasahara, T. Terashima, T. Shibauchi, 
Y. Matsuda, A. Chainani, and S. Shin,
Sccience {\bf 332}, 564 (2011).

\bibitem{Yoshida}
T. Yoshida, S. Ideta, T. Shimojima, W. Malaeb, K. Shinada, H. Suzuki, I. Nishi, A. Fujimori, K. Ishizaka, S. Shin, Y. Nakashima, H. Anzai, M. Arita, A. Ino, H. Namatame, M. Taniguchi, H. Kumigashira, K. Ono, S. Kasahara, T. Shibauchi, T. Terashima, Y. Matsuda, M. Nakajima, S. Uchida, Y. Tomioka, T. Ito, K. Kihou, C. H. Lee, A. Iyo, H. Eisaki, H. Ikeda, R. Arita, T. Saito, S. Onari, and H. Kontani,  arXiv:1301.4818.


\bibitem{Yamashita}
M. Yamashita, Y. Senshu, T. Shibauchi, S. Kasahara, K. Hashimoto, D. Watanabe, H. Ikeda, T. Terashima,
I. Vekhter, A. B. Vorontsov, and Y. Matsuda, Phys. Rev. B {\bf 84}, 060507(R) (2011).



\bibitem{Takimoto}
T. Takimoto, T. Hotta, T. Maehira and K. Ueda,
J. Phys. Condens. Matter {\bf 14}, L369 (2002).

\bibitem{Scalapino}
S. Graser, A. F. Kemper, T. A. Maier, H.-P. Cheng, P. J. Hirschfeld, and D. J. Scalapino,
New Journal of Physics. {\bf 12}, 073030 (2010).

\bibitem{Chubukov-spmh}
S. Maiti and A. V. Chubukov, Phys. Rev. B {\bf 87}, 144511 (2013).

\bibitem{Ahn}
F. Ahn, I. Eremin, J. Knolle, V.B. Zabolotnyy, S.V. Borisenko, 
B. B\"{u}chner, A.V. Chubukov, arXiv:1402.2112.
\bibitem{Yin}
Z. P. Yin, K. Haule, G. Kotliar, arXiv:1311.1188.

\bibitem{Kontani-Raman}
H. Kontani and Y. Yamakawa,  arXiv:1312.0528

\bibitem{Onari-SCVCS}
S. Onari, H. Kontani, S. V. Borisenko, V.B. Zabolotnyy and  B. B\"{u}chner,
arXiv:1307.6119.

\bibitem{Onari-Hdoped}
S. Onari, Y. Yamakawa and H. Kontani, arXiv:1312.0481
(Phys. Rev. Lett. (2014)).

\bibitem{Ohno-SCVC}
Y. Ohno, M. Tsuchiizu, S. Onari, and H. Kontani, J. Phys. Soc. Jpn. 
{\bf 82}, 013707 (2013). 

\bibitem{Tsuchiizu}
M. Tsuchiizu, Y. Ohno, S. Onari, and H. Kontani, 
Phys. Rev. Lett. {\bf 111}, 057003 (2013).

\bibitem{comment}
In Ref. \cite{Onari-SCVCS},
the quadrupole susceptibility
$\chi_{yz}(\q)\propto \chi_{2,4;4,2}(\q)$ given by the SC-VC$_\Sigma$ method
has commensurate peak at $\q=(\pi,0)$,
since the calculation temperature is rather high.

\bibitem{DMFT}
Z. Yin, K. Haule, and G. Kotliar, Nat. Mater. {\bf 10}, 932
(2011);
J. Ferber, K. Foyevtsova, R. Valenti, and H. O. Jeschke,
Phys. Rev. B {\bf 85}, 094505 (2012);
G. Lee {\it et al.}, Phys. Rev. Lett. {\bf 109}, 177001 (2012).

\bibitem{dHvA1}
C. Putzke, et al., Phys. Rev. Lett. {\bf 108}, 047002 (2012).

\bibitem{dHvA2}
B. Zeng, et al., Phys. Rev. B {\bf 88}, 144518 (2013).

\bibitem{Platt}
C. Platt, R. Thomale, and W. Hanke, 
Phys. Rev. B {\bf 84}, 235121 (2011).



\end{thebibliography}
\end{document}